\documentclass{emulateapj}
\usepackage{apjfonts}
\usepackage{amsmath}

\usepackage{booktabs}
\usepackage{color}
\usepackage{graphicx}
\usepackage{amsmath}

\begin{document}

\title{Magnetic Field Saturation of the Ion Weibel Instability in Interpenetrating Relativistic Plasmas}

\author{Makoto Takamoto}
\affil{Department of Earth and Planetary Science, The University of Tokyo, Hongo, Bunkyo-ku, Tokyo 113-0033, Japan}

\author{Yosuke Matsumoto}
\affiliation{Department of Physics, Chiba University 1-33 Yayoi-cho, Inage-ku, Chiba 263-8522, Japan}

\author{Tsunehiko N. Kato}
\affiliation{Center for Computational Astrophysics, National Astronomical Observatory of Japan, 2-21-1 Osawa, Mitaka, Tokyo 181-8588, Japan}



\begin{abstract}
The time evolution and saturation of the Weibel instability at the ion Alfv\'en current are presented by ab initio particle-in-cell simulations. We found that the ion Weibel current in 3D could evolve into the Alfv\'en current where the magnetic field energy is sustained at 1.5\% of the initial beam kinetic energy. The current filaments are no longer isolated at saturation, but rather connected to each other to form a network structure. Electrons are continuously heated during the coalescence of the filaments, which is crucial for obtaining sustained magnetic fields with much stronger levels than with 2D simulations. The results highlight again the importance of the Weibel instability in generating magnetic fields in laboratory, astrophysical, and cosmological situations.  
\end{abstract}



\section{Introduction} \label{sec:intro}
The magnetic field is a fundamental physical quantity found everywhere in the universe. It plays essential roles in galaxy spiral arms \citep{fletcher2011}, high-energy astrophysical phenomena \citep{1976MNRAS.175..613S,1999PhR...314..575P,2014RPPh...77f6901B}, star formation \citep{1994ApJ...432..720B,2008ApJ...676.1088M}, and cyclic dynamos at the interior of stars and planets \citep{glatzmaier1999,2016Sci...351.1427H}. Because of its ubiquity, the origin of the magnetic field is of great interest in various research fields.

The Weibel instability arises from anisotropy in the plasma velocity distribution function \citep{1959PhRvL...2...83W,1959PhFl....2..337F}. Because the instability feeds the free energy of the anisotropy and converts it into magnetic energy,  it is one of the promising mechanisms responsible for the origin of the magnetic field. Following early theoretical works \citep{1999ApJ...526..697M,2004ApJ...608L..13F,2004ApJ...617L.107H,2005PhPl...12h0705K,2006ApJ...642L...1M,2006AIPC..856..109S,2007AA...475...19A,2007AA...475....1A,2009ApJ...699..990B}, numerical simulations have revealed its wide applicability. The generated magnetic field plays crucial roles in collisionless shock formation \citep{2007ApJ...668..974K,2008ApJ...673L..39S,2008ApJ...681L..93K}, associated particle accelerations \citep{sironi2013,2015Sci...347..974M}, and afterglows of gamma-ray bursts \citep{2006ApJ...642..389N,tomita2016}. 
Recent experiments using high power laser facilities have succeeded in showing detailed images of the nonlinear structure characterized by filaments of the current \citep{2013PhRvL.111v5002F,2015NatPh..11..173H,2015PhPl...22e6311P,2017PhPl...24d1410H}. 

\citet{2005PhPl...12h0705K} proposed a theoretical model describing magnetic field evolution in cylindrical currents of the electron Weibel instability, and found that the maximum magnetic field could be obtained when the current reached the so-called \textit{Alfv\'en current} \citep{1939PhRv...55..425A,Lawson1958,Honda2000}, 
\begin{equation}
I_{\rm A,s} = I_0 \langle \gamma \beta_{||} \rangle,
\label{eqn:alfven}
\end{equation}
where $I_0 \equiv m_{s} c^3/q_{s}$, $m_{s}$ and $q_{s}$ are the mass and charge of the particle of each species $s$, respectively, and $c$ is speed of light. $\beta_{||}$ is the magnitude of the three-velocity in the unit of light speed along the beam direction, and $\gamma$ is the Lorentz factor. The angle brackets $\langle \rangle$ denote an average over the beam volume. This gives an upper limit of the cylindrical current in which the Larmor radius of particles in the self-generated magnetic field is comparable with the filament size, and thus determines the magnetic field saturation.

The saturation mechanism of the ion Weibel instability is not well understood despite its importance in understanding attainable maximum magnetic energy. Using particle-in-cell (PIC) simulations, \citet{2004ApJ...608L..13F} reported early evolution of the ion Weibel filaments and found that the generated magnetic field keeps growing even after saturation of the electron Weibel instability. In contrast, \citet{2015ApJ...806..165K} reported that the Weibel filaments finally disappear in the two-dimensional (2D) case because of strong electric fields generated by either the Buneman instability or by an oblique unstable mode \citep{2009ApJ...699..990B}. More recently, an analytical model of coalescence of the ion Weibel filaments was proposed for non-relativistic plasmas based on the kinetic quasi-linear theory of the instability, and was validated by 2D and 3D PIC simulations in the weak nonlinear regime \citep{Ruyer2015,2016PhRvL.117f5001R}.

In this Letter, we present the evolution and saturation of the Weibel instability in interpenetrating relativistic ion-electron plasmas by means of three-dimensional (3D) PIC simulations. Large-scale, long-term simulations enabled us to elucidate for the first time the coalescence of ion-scale current filaments and the resulting magnetic field saturation after reaching the ion Alfv\'en current limit.

\section{Numerical Setups} \label{sec:setups}
To explore the nonlinear evolution of the Weibel instability, we used a fully kinetic electromagnetic PIC code \citep{matsumoto2017} which implements the quadratic function for the particle weighting and the implicit FDTD scheme for the Maxwell equation. Simulation runs were initialized by two cold unmagnetized counter flows with a bulk Lorentz factor of $\gamma = 5$ in the laboratory frame. The beams were set in the $x$-direction with the periodic boundary condition in the all directions. The cold counter-streaming condition induced a very large anisotropy of the velocity distribution in the system. Consequently, the relativistic Weibel instability grew rapidly in the whole region of the simulation domain. In the following, time and spatial length were normalized to the inverse of the non-relativistic ion plasma frequency $\omega_{p,i}^{-1}$ and the ion inertial length $c/\omega_{p,i}$ using the initial values, respectively. Ten particles per cell were used for each species of ion and electron with the mass ratio of $M/m=25$\footnote{We have confirmed that present results and conclusions are similarly obtained when we examined with different numbers of particles per cell. The total energy fluctuates within $\pm 0.85\%$ errors in the present long-term simulations.}. The cell size was set as $\Delta = 0.1\  c/\omega_{p,e}$, equivalent to $0.045\  c/\tilde{\omega}_{p,e}$ in terms of the proper plasma frequency $\tilde{\omega}_{p,e}$ to resolve the initial electron Weibel instability accurately. The numerical Cherenkov instability in the relativistic plasma flows was suppressed by carefully choosing a magical time step size of $\Delta t = \Delta/c$ ($=0.1\omega_{pe}^{-1}$) specific to our semi-implicit PIC code \citep{2015PASJ...67...64I}. The simulation domain size was taken as $L_{\rm x} \times L_{\rm y} \times L_{\rm z} = 20.0\ c/\omega_{p,i} \times 51.2\ c/\omega_{p,i} \times 51.2\ c/\omega_{p,i}$. Such very large-scale PIC simulations were performed on the Japanese K computer using $16,384$ nodes ($131,072$ processor cores) and $200$ TB of physical memory. 

\section{Temporal Evolution} \label{sec:evolutions}
\begin{figure}[t]
 \centering
  \includegraphics[width=0.475\textwidth]{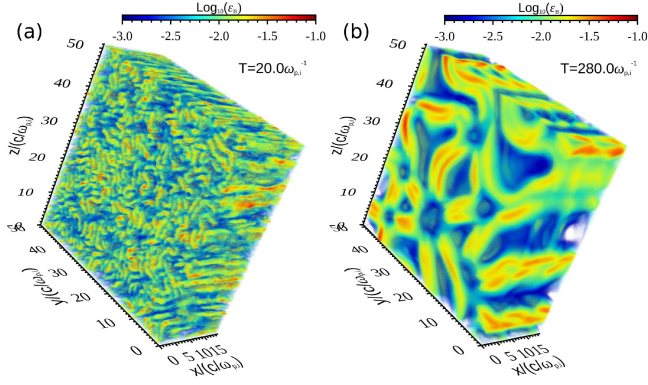}
  \caption{3D profiles of the magnetization parameter $\epsilon_{\rm B}$ color-coded in a logarithmic scale at characteristic time stages of (a) the electron Alfv\'en current ($t=20\ \omega_{p,i}^{-1}$), and (b) the ion Alfv\'en current ($t=280\ \omega_{p,i}^{-1}$).}
  \label{fig:3.0}
\end{figure}
Figure \ref{fig:3.0} shows 3D profiles of the magnetization parameter $\epsilon_{\rm B} \equiv B^2/8\pi M n_0 c^2 (\gamma - 1)$, where $B$ is the magnetic field strength, representing the ratio of the generated magnetic energy to the initial beam kinetic energy. At a characteristic time stage of the electron Alfv\'en current ($t=20\ \omega_{p,i}^{-1}$), the simulation domain is filled with filamentary structures elongated in the background beam direction with a typical size of 10 times the electron inertia length (Fig. \ref{fig:3.0}(a)). These filaments continue to coalesce into larger scales (inverse cascade) beyond a time stage of $t=280\ \omega_{p,i}^{-1}$ at which the ion Alfv\'en current is reached (Fig. \ref{fig:3.0}(b)). 

\begin{figure}[t]
 \centering
  \includegraphics[width=0.4\textwidth]{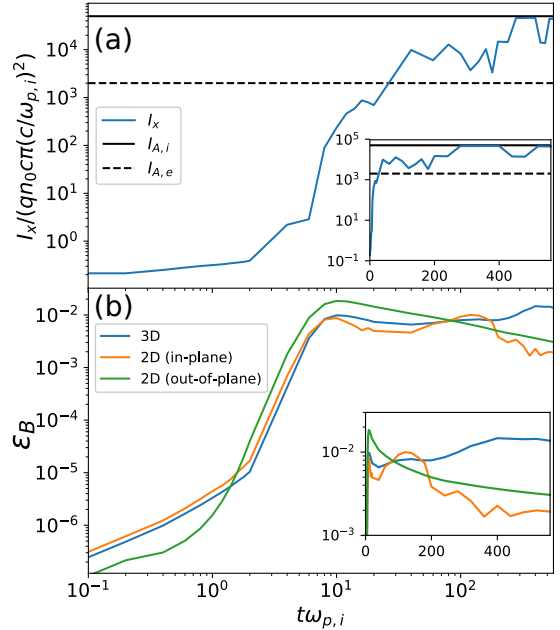}
  \caption{(a) Temporal evolution of the mean current in a filament (blue) along with characteristic currents of the electron Alfv\'en current (dashed black) and the ion Alfv\'en current (solid black).
           (b) Temporal evolution of the averaged magnetization parameter for 3D (blue), 2D in-plane (orange), and 2D out-of-plane (green) cases. Insets are the same format but with the abscissa on the linear scale in time to emphasize the later evolution.
           }
  \label{fig:3.1.0}
\end{figure}
Temporal evolution of the mean current in a filament is shown in Figure \ref{fig:3.1.0}(a), and is compared with the Alfv\'en current  (eq. (\ref{eqn:alfven})) for each species of the ion ($s=i$) and electron ($s=e$) using the initial beam speed. Here, the current is $\displaystyle I = \langle|J|\rangle_{yz} \pi R^2$, where $J$ is the current density and $R$ was measured as a half wavelength at the peak of the Fourier-transformed current density in the $y$--$z$ plane, and $\langle\rangle_{yz}$ denotes the area average in the $y$--$z$ plane. The current reached the electron Alfv\'en current at $t=20\ \omega_{p,i}^{-1}$, after which it continuously grew and finally reached the ion Alfv\'en current at $t=280\ \omega_{p,i}^{-1}$. The current did not exceed the Alfv\'en current limit up to $560\ \omega_{p,i}^{-1}$ in the present simulation run.

The magnetic field energy ($\epsilon_{\rm B}$ averaged in the simulation domain) also saturated around the time when the current reached the ion Alfv\'en current and sustained at $\epsilon_{\rm B} \sim 0.015$ (Fig. \ref{fig:3.1.0}(b)). This strong magnetic field level was maintained until the simulation run ended at $t=560\ \omega_{p,i}^{-1}$. Comparisons with 2D simulations exhibited distinct differences in the saturation phase even using the same parameters. In the 2D in-plane current case (simulations in the $x$--$z$ plane), $\epsilon_{\rm B}$ evolved similarly to the 3D case until $t=100\ \omega_{p,i}^{-1}$. After that, however, $\epsilon_{\rm B}$ decreased rapidly. In the 2D out-of-plane current case (simulations in the $y$--$z$ plane), $\epsilon_{\rm B}$ saturated at $t = 10\  \omega_{p,i}^{-1}$, and then $\epsilon_{\rm B}$ decreased monotonically. 

Regarding particle acceleration, we did not observe non-thermal particles clearly during the saturation process.
The particle velocity distribution functions for protons and electrons are both essentially thermal.

\section{3D vs. 2D (in-plane)} \label{sec:inplane}
\begin{figure}[t]
 \centering
 \includegraphics[width=0.375\textwidth]{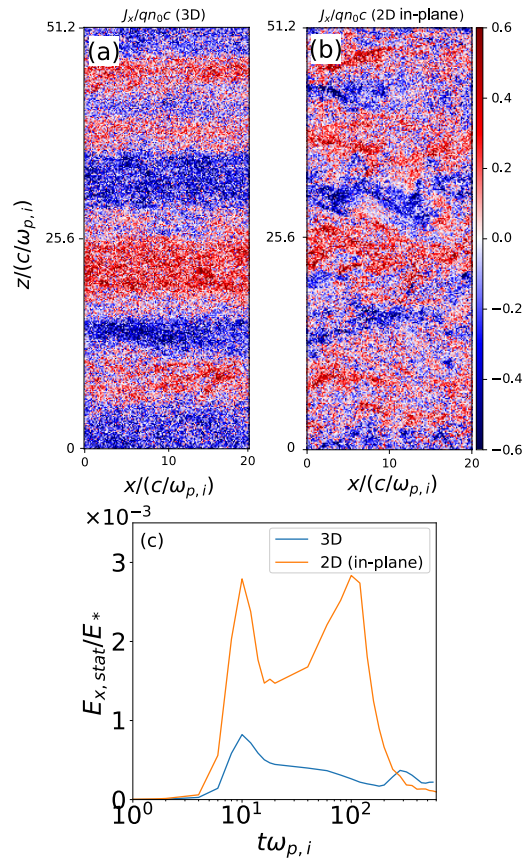}
  \caption{2D profile of the $x$-component of the current density at $t=60\ \omega_{p,i}^{-1}$ for (a) 3D and (b) 2D in-plane cases. (c) Temporal evolution of the $x$-component of the electrostatic field strength in the unit of $E_* \equiv (4 \pi M n_{\rm i} c^2)^{1/2}$.}
  \label{fig:3.1.1}
\end{figure}
Figures \ref{fig:3.1.1}(a) and \ref{fig:3.1.1}(b) present 2D profiles of the current density for 3D and 2D in-plane cases. It clearly shows that the current filaments in the 2D in-plane case are much more fragmented than in the 3D case, whereas in the 3D case, fluctuations between filaments can be found. The difference stems from the geometrical restriction in the 2D in-plane case, in which a current filament first collides with a filament with the different polarity in the course of coalescence with the nearby filament with the same polarity. This results in enhanced electrostatic field excitation by the beam instability \citep{2015ApJ...806..165K}. Indeed, the breakup of the filaments in the case of 2D in-plane occurred right after the peak of the electrostatic field ($t=10\ \omega_{p,i}^{-1}$) as shown in Figure \ref{fig:3.1.1}(c). In contrast in the 3D case, oppositely signed filaments can pass each other without collisions during the coalescence of the filaments with the same polarity, resulting in weak excitation of the electrostatic mode (blue curve in Fig. \ref{fig:3.1.1}(c)). The results suggest that the breakup of current filaments and thus rapid decay in the magnetic energy are characteristic only in the 2D (in-plane) case \citep{2011ApJ...739L..42H}.

\section{3D vs. 2D (out-of-plane)} \label{sec:outofplane}
\begin{figure}[t]
 \centering
 \includegraphics[width=0.475\textwidth]{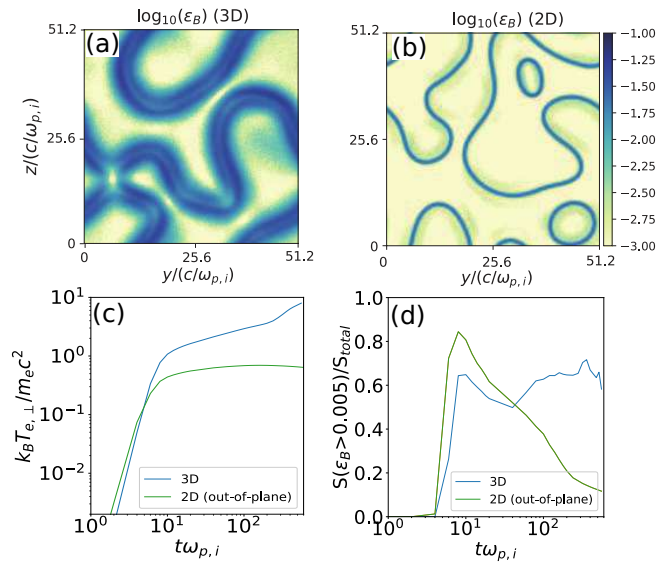}
  \caption{Snapshot of $\epsilon_{\rm B}$ in the $y$--$z$ plane at $t=560\ \omega_{p,i}^{-1}$ for (a) 3D and (b) 2D out-of-plane cases in a logarithmic scale. 
(c) Temporal evolution of the area where $\epsilon_{\rm B} > 5 \times 10^{-3}$ for the 3D (blue) and 2D out-of-plane (green) cases.
(d) Temporal evolution of the electron temperature perpendicular to the initial flow direction.}
  \label{fig:3.1.3}
\end{figure}
Geometrically, it could be expected that the current filaments in the 2D out-of-plane case would evolve similarly to the 3D case because the current filaments can avoid collision with the neighboring oppositely flowing ones.  Figure \ref{fig:3.1.0}(b), however, shows that the magnetic energy monotonically decreased after saturation at $t=10\ \omega_{p,i}^{-1}$, and finally became much smaller than in the 3D case. Figure \ref{fig:3.1.3} shows snapshots of $\epsilon_{\rm B}$ for the 3D (Fig. \ref{fig:3.1.3}(a)) and the 2D out-of-plane (Fig. \ref{fig:3.1.3}(b)) cases in the final evolution stage at $t=560\ \omega_{p,i}^{-1}$. In the late nonlinear stage, the currents were no longer isolated cylindrical beams \citep{2005PhPl...12h0705K}, but were rather connected to each other to form a network structure. It was found that strong magnetic field areas in the 3D case were distributed broadly at the interface of the currents whereas in the 2D out-of-plane case they were confined in a very narrow layer. The local maximum values, however, were similar to each other ($\epsilon_{\rm B} \sim 0.1$).

The broad strong magnetic field areas in the 3D case can be attributed to an electron gyro motion around the current interface in the nonlinear stage. Figure \ref{fig:3.1.3}(c) shows the temporal evolution of the electron temperature perpendicular to the initial beam direction ($x$-direction) for the 3D (blue curve) and the 2D out-of-plane (green curve) cases. We found in the 3D result that electrons were heated preferentially in the perpendicular direction during coalescence of ion-scale filaments, and were finally energized to the relativistic temperature ($k_B T_\perp \sim 10\ mc^2$) at $t=560\ \omega_{p,i}^{-1}$, whereas the 2D result showed a heating process only in the very early stage, followed by a constant nonrelativistic temperature ($k_B T_\perp \sim 0.5\ mc^2$).

In the 3D case, the non-uniform structures in the $x$-direction (Fig. \ref{fig:3.1.1}(a)) can cause mixing of current filaments in the $x$-direction, resulting in significant heating of the electrons. This allows the broadening of the strong magnetic field area, and hence the total $\epsilon_{\rm B}$ is 10 times larger than the 2D out-of-plane case. 
Figure \ref{fig:3.1.3}(d) is the temporal evolution of the area where $\epsilon_{\rm B} > 5 \times 10^{-3}$, that is, the strongly magnetized region. In the 2D out-of-plane case, the strongly magnetized  area quickly decreased after its peak in the early stage at $t=10\ \omega_{p,i}^{-1}$, whereas in the 3D case, it was sustained even in the late nonlinear stage. Note that the evolution of the strongly magnetized region in Figure \ref{fig:3.1.3}(d) coincides with Figure \ref{fig:3.1.0}(b), indicating that the different evolution of $\epsilon_{\rm B}$ between the 3D and the 2D out-of-plane cases was the result of different filling factors of strongly magnetized regions in the simulation domain rather than the magnetic field strength itself.

\section{Summary and Discussion} \label{sec:summary}
We have presented the saturation mechanisms of the ion Weibel instability by means of long-term, large-scale PIC simulations. It was found that the Weibel-generated magnetic fields sustained for long time periods after reaching the ion Alfv\'en current limit in the 3D space; the Weibel filaments are stable during the filament merging process against secondary instabilities such as the Buneman and other electrostatic modes found in 2D in-plane simulation studies. We found that electron heating processes continue during coalescence of ion-scale filaments in the late nonlinear stage and are crucial for the sustained magnetic field. 
Concerning the afterglows of gamma-ray bursts, 
the obtained value, $\epsilon_{\rm B} \sim 0.015$, supports a model with moderately strong magnetic field \citep{1999ApJ...526..697M}. 
More importantly, our study showed that the magnetic field generated by Weibel instability is stably sustained much longer than the plasma timescale, that is, at least for 560 ion plasma oscillations. 
Note that some recent study of the timescale of the afterglow emission in the downstream reported the following estimation~\citep{2007ApJ...671.1858S,tomita2016}, $7.4 \times 10^7 \ \omega_{\rm p,i}^{-1} \ E_{\rm iso, 53}^{1/3} n_{\rm 0}^{1/6} \Gamma_{100}^{-5/3}$ 
where $E_{\rm iso, 53}$ is the isotropic equivalent energy normalized to $10^{53} \mathrm{erg}$,
$n_0$ is the number density of interstellar medium normalized to $1 \mathrm {cm}^{-3}$,
and $\Gamma_{100}$ is the Lorentz factor of the blastwave normalized to $100$.
Although the simulation timescale is still much shorter than this one, our results imply a possibility of a much more stable Weibel magnetic field than previously thought. 

The present results indicate that collisionless shock simulations in 2D in-plane configurations underestimated the shock forming time, and overestimated the particle acceleration efficiency by the electrostatic field in the shock transition region. This is because the Buneman type instability observed in the 2D in-plane case was suppressed in our 3D results, and the Weibel instability becomes dominant in relativistic and weakly magnetized counter flows \citep{2009ApJ...699..990B}. Stochastic particle accelerations with Weibel-generated magnetic turbulence \citep{2015SSRv..191..519S} are more important than the direct acceleration by the electric field.
As shown in the present work, the magnetic turbulence generated by the Weibel instability can be stronger in 3D than in 2D.
Thus, this acceleration process can be more efficient in 3D than those observed
in 2D simulations so far \citep{matsumoto2017}.

Finally, 
we consider that the present results can 
provide a guideline for designing laboratory experiments using high-power laser facilities aiming for understanding magnetic field generation by the Weibel instability and collisionless shock formation. The experiment time scale of $\sim 200\ \omega_{p,i}^{-1}$ and the spatial scale size of $20\ c/\omega_{p,i} \times 50\ c/\omega_{p,i} \times 50\ c/\omega_{p,i}$ were required to allow current filaments to evolve into the ion Alfv\'en current and true magnetic field saturation. 

\acknowledgments

We would like to thank T. Amano, M. Hoshino, and Y. Ohira for many fruitful comments and discussions. 
This research used computational resources of the K computer provided by the RIKEN Advanced Institute for Computational Science through the HPCI System Research project (Project ID:hp160121, hp170125, hp170231) and Cray XC30 at   Center for Computational Astrophysics, National Astronomical Observatory of Japan. This work was supported in part by the Postdoctoral Fellowships by the Japan Society for the Promotion of Science No. 201506571 (M. T.), and JSPS KAKENHI Grant Number 17H02877 (Y. M.).




\end{document}